\documentclass{article}
\usepackage{spconf,amsmath,graphicx}

\usepackage[rgb]{xcolor}
\usepackage{hyperref}
\hypersetup{
    colorlinks=true,
    linkcolor=blue,
    filecolor=magenta,
    urlcolor=cyan,
    citecolor=blue,
}
\usepackage{colortbl}

\usepackage{cite}
\usepackage{amsmath}
\usepackage{url}
\usepackage{amsfonts}
\usepackage{multirow}
\usepackage{makecell}
\usepackage{booktabs} 
\usepackage{textcomp}

\usepackage{pifont}

\usepackage{subcaption}

\def\H{{\mathbf{H}}}
\def\Asoft{{\mathcal{A}_{\text{soft}}}}
\def\Ahard{{\mathcal{A}_{\text{hard}}}}

\title{AAS-VC: On the Generalization Ability of Automatic Alignment Search based Non-autoregressive Sequence-to-sequence Voice Conversion}

\name{\begin{tabular}{c}
    \it Wen-Chin Huang$^1$, Kazuhiro Kobayashi$^{1,2}$, Tomoki Toda$^1$
\end{tabular}}

\address{$^{1}$Nagoya University, Japan \\
  $^{2}$TARVO, Inc., Japan}

\begin{document}
\ninept
\maketitle

\begin{abstract}
Non-autoregressive (non-AR) sequence-to-seqeunce (seq2seq) models for voice conversion (VC) is attractive in its ability to effectively model the temporal structure while enjoying boosted intelligibility and fast inference thanks to non-AR modeling. However, the dependency of current non-AR seq2seq VC models on ground truth durations extracted from an external AR model greatly limits its generalization ability to smaller training datasets. In this paper, we first demonstrate the above-mentioned problem by varying the training data size. Then, we present AAS-VC, a non-AR seq2seq VC model based on automatic alignment search (AAS), which removes the dependency on external durations and serves as a proper inductive bias to provide the required generalization ability for small datasets. Experimental results show that AAS-VC can generalize better to a training dataset of only 5 minutes. We also conducted ablation studies to justify several model design choices. The audio samples and implementation are available online.
\end{abstract}

\section{Introduction}
\label{sec:intro}

Voice conversion (VC) has seen rapid improvement over the past few decades thanks to the advance in neural waveform generation and content disentanglement \cite{vcc2020, vc-survey-2021}. Auto-encoder style training based on the recognition-synthesis framework has become the de facto in VC as it allows training based on non-parallel corpora and even unlocks zero-shot VC based on a single utterance from the target speaker. One pitfall of such a framework is the inability to well model the temporal structure, which greatly influences speaking style and thus speaker similarity. To achieve such a goal, it is still considered the most effective to use paired utterances from the source and target. In many real-world VC problems such as accent conversion, dysarthric VC and eletrolaryngeal (EL) speech enhancement, it is often acceptable to relax the datast constraint and assume that a parallel dataset is available, although its size is often limited in practice. Under such a setting, sequence-to-sequence (seq2seq) is considered the standard choice.

Seq2seq modeling aims to find the mapping from the input to the output sequence of arbitrary lengths, and has been widely used in speech and natural language processing. While autoregressive (AR) seq2seq models have been successfully applied to VC \cite{VTN} , the time-consuming step-by-step sampling nature hinders it from real-time processing, which is crucial for practical VC applications. Inspired by text-to-speech (TTS) models like FastSpeech2 \cite{fastspeech2}, non-autoregressive (non-AR) seq2seq modeling has been successfully applied to VC \cite{cfs2-vc}, which we will refer to as FS2-VC. 
The key to success of FS2-VC was to separate acoustic feature generation and alignment (essentially, duration) modeling by ``outsourcing'' the latter to an external pre-trained AR seq2seq model, and as a result outperforming AR seq2seq VC models in terms of generation speed and quality. However, it is not hard to see the disadvantages of FS2-VC: in addition to the complicated training pipeline, a more serious problem is that the non-AR model training depends on the quality of the durations generated by the AR model. This is especially crucial in low-resource settings, where the alignment from the AR model itself is already unstable.

It is therefore of significant interest to jointly perform alignment and acoustic modeling in non-AR seq2seq VC, while maintaining the fast inference speed and high-quality. A recent attempt was made in \cite{e2e-s2s-vc} by directly modifying two modern end-to-end non-AR TTS models, VITS \cite{vits} and JETS \cite{jets}, for VC. These models imposed alignment motononicity during training, as it served as a strong inductive bias to ensure effective alignment extraction. However, the resulting models, namely VITS-VC and JETS-VC, are much more complicated compared to FS2-VC, as they include techniques like adversarial learning, variational inference and flow-based probabilistic modeling. With a very limited amount of ablation studies, there are much left to investigate in such a framework.

In this work, we focus on the effectiveness of the alignment learning module, especially on its generalization ability to reduced training data size and unseen source speakers. By starting from FS2-VC, we maintain the parts related to acoustic modeling, and make minimal changes to incorporate the \textbf{a}utomatic \textbf{a}lignment \textbf{s}earch (AAS) module. The resulting model, named \textbf{AAS-VC}, is trained on a simple L1 loss in the target acoustic feature domain, a loss for the duration predictor, and two losses for alignment learning.
Through experimental evaluations, we first conduct ablation studies to justify the model designs of AAS-VC. We then demonstrate the following main contributions of this work:
\begin{itemize}
    \item The performance of FS2-VC depends on the quality of the alignments from the AR model, especially in a low-resource setting (i.e. 5 min of training data).
    \item AAS-VC can almost perform on par with FS2-VC in a high-resource setting, and can generalize better in a low-resource setting.
\end{itemize}

\vspace{-10pt}
\section{Non-AR seq2seq VC framework}
\label{sec:nar-seq2seq-vc}

Instead of directly applying existing non-AR TTS models with AAS ability, the philosophy in this work is to start from the non-AR seq2seq VC framework proposed in \cite{cfs2-vc}, and make minimal changes to examine how AAS influences the non-AR seq2seq VC model. Thus, we start by describing a general non-AR seq2seq VC framework, which is based on that proposed in \cite{cfs2-vc} but we keep the most essential parts and leave flexibility for possible improvements changes.

\begin{figure}[t]
	\centering
	\includegraphics[width=0.9\linewidth]{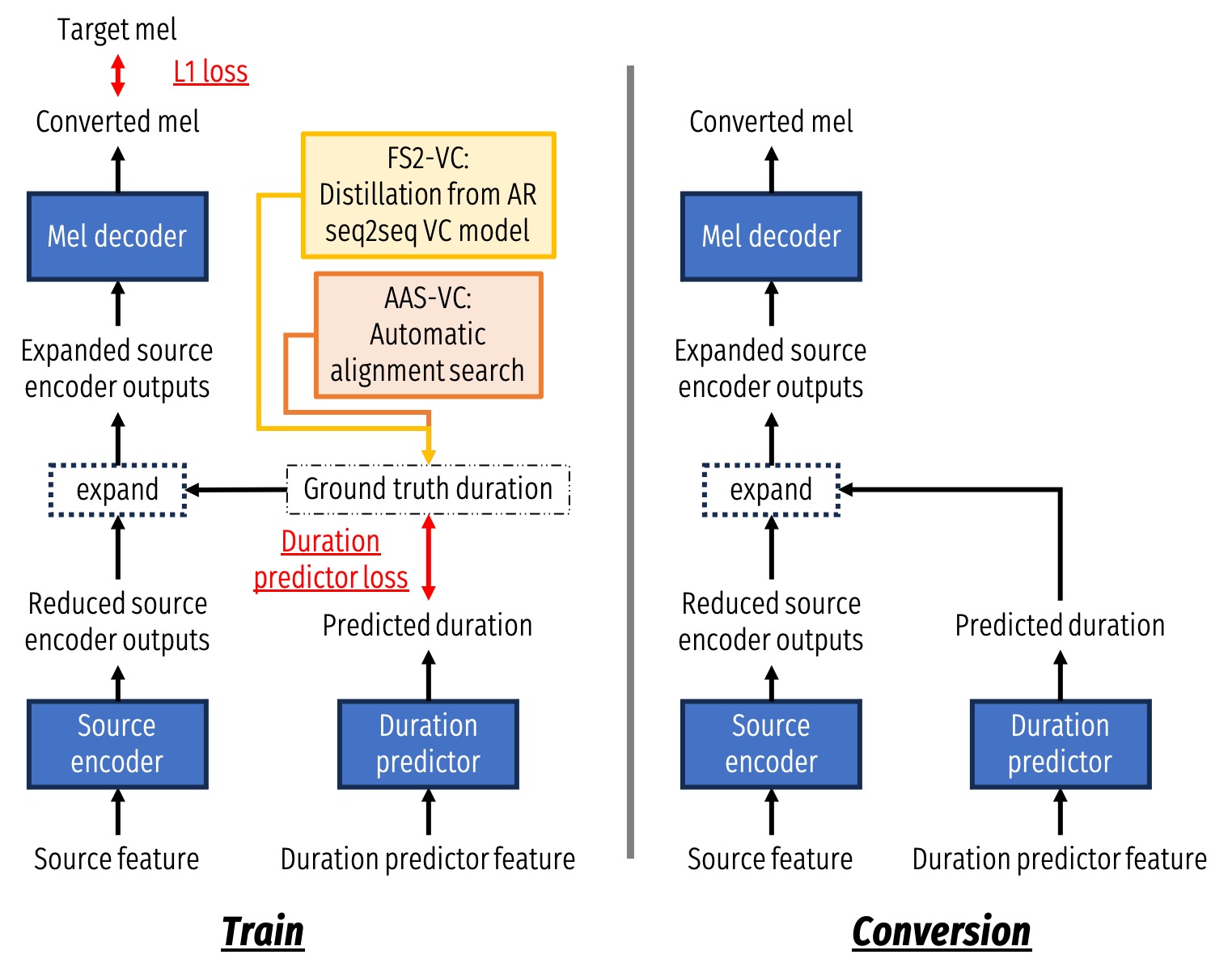}
	\caption{\label{fig:overview}Training and conversion processes of the non-autoregressive sequence-to-sequence voice conversion framework in this work.}	
        \vspace{-5pt}
\end{figure}
\begin{figure}[t]
	\centering
	\includegraphics[width=0.9\linewidth]{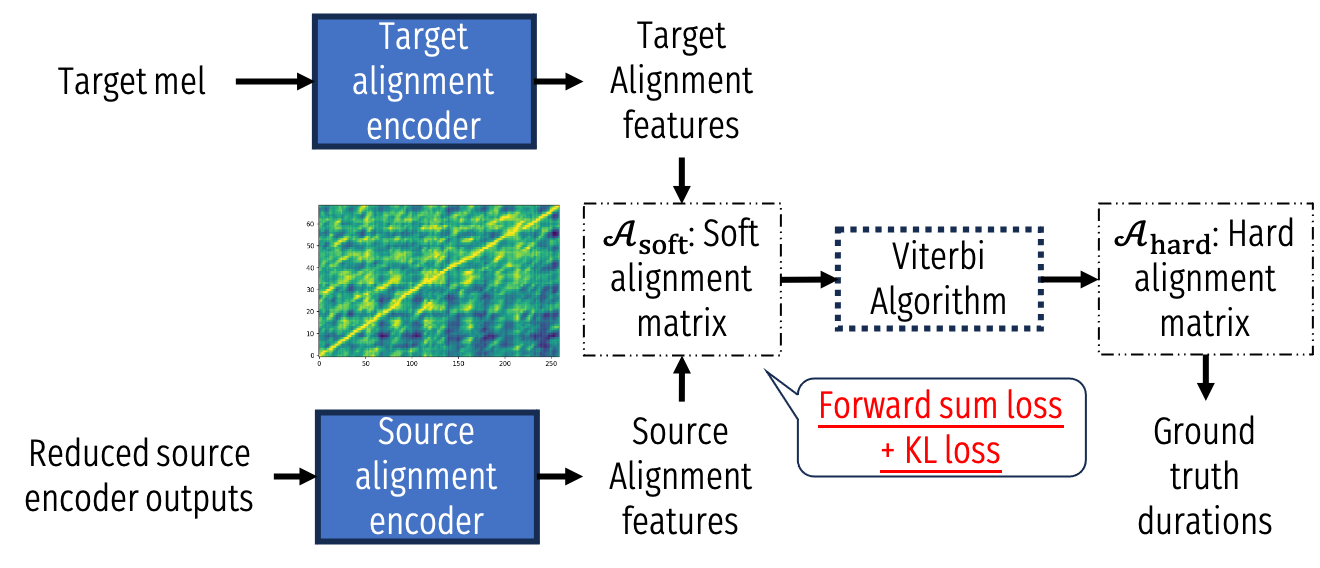}
	\caption{\label{fig:aas}Automatic alignment search.}
        \vspace{-5pt}
\end{figure}

\subsection{Overview}

Figure~\ref{fig:overview} shows the overview of the non-AR seq2seq VC framework in this work. The training dataset is a set of parallel utterances from the source and the target. While we extract mel-spectrograms from the target waveforms as the training target in this work, the source features is not necessary the same and we will discuss the feature choice in Section~\ref{ssec:feature}. Unless specified, all modules are consisted of Conformer blocks \cite{conformer}, as they exploit both local and global features well in speech processing tasks \cite{espnet-conformer}.

During training, the source features first go through a source encoder. The length of the source encoder output sequence is often reduced.
Then, the source encoder output feature at each time steps is allocated an integer called the ground truth duration, and an ``expand''\footnote{Note that despite of the naming ``expand'', the allocated duration can be 0 such that the feature is simply removed in the succeeding operations.}\footnote{In many literature, this operation is also called ``length regulation''.} operation adjusts the length of the sequence. Meanwhile, a duration predictor conditioned on features extracted from the source utterance is trained to match the ground truth duration sequence, and the optimization is independent with the rest of the modules. Finally, a mel decoder synthesizes the converted mel-spectrogram from the expanded source encoder outputs, and a vocoder generates the final waveform. The training objective $\mathcal{L}_{\text{NAR}}$ includes a L1 loss $\mathcal{L}_1$ between the converted and target mel-spectrograms, as well as a duration prediction loss $\mathcal{L}_{\text{DP}}$:

\begin{equation}
    \mathcal{L}_{\text{NAR}}=\mathcal{L}_1 + \mathcal{L}_{\text{DP}}. \label{eq:nar-vc}
\end{equation}

The conversion process is nearly identical to that of training, except that instead of the ground truth durations, the predicted durations from the duration predictor are used in the ``expand'' operation.

\vspace{-5pt}
\subsection{FS2-VC: duration distillation from an AR seq2seq model}
\label{ssec:fs2-vc}

In FS2-VC \cite{cfs2-vc}, the ground truth duration is distilled from an AR seq2seq VC model trained on the same dataset. Specifically, the durations are calculated from the attention weights obtained by running the forward process in teacher forcing mode, which assures that the summation of the durations matches that of the target mel-spectrogram. As stated in Section~\ref{sec:intro}, duration distillation complicates the training pipeline, since an AR model needs to be trained beforehand. Furthermore, as AR models are known to be prone to attention errors, the training of both the mel decoder and the duration predictor will be affected by the erroneous durations. Especially, when only a very small parallel dataset is available, the durations should suffer from such errors even more seriously.

\section{AAS-VC: Non-AR seq2seq VC with automatic alignment search}

The AAS-VC framework is based on the framework illustrated in Figure~\ref{fig:overview}, with the biggest difference being how the ground truth duration sequence is generated directly from the source and target utterances alone, without the help of any external module.

\subsection{Automatic alignment search}
\label{ssec:aas}

The automatic alignment search process, illustrated in Figure~\ref{fig:aas}, is mostly based on that proposed in \cite{rad-tts} and then adopted in VITS-VC and JETS-VC \cite{e2e-s2s-vc}. The core idea is to apply monotonicity constraints as a strong inductive bias. First, the two input features from the source and target, namely the reduced source encoder output and the target mel-spectrogram, are passed through their respectively alignment encoders. The idea of the alignment search is that feature frames with similar contents should be aligned. Thus, by denoting the source and target alignment features as $\H^{\text{src}}=\{{h^{\text{src}}_1, \cdots, h^{\text{src}}_{T_{\text{src}}}}\}$ and $\H^{\text{trg}}=\{{h^{\text{trg}}_1, \cdots, h^{\text{trg}}_{T_{\text{trg}}}}\}$, respectively, the soft alignment matrix $\Asoft$ is calculated by applying the softmax function over each column in a distance matrix $\mathcal{D}$. The $(i, j)$-th elements of $\Asoft$ and $\mathcal{D}$ are given by:
\begin{align}
    (\mathcal{D})_{i,j} &= \lVert h^{\text{src}}_i - h^{\text{trg}}_j \rVert_2,\\
    (\Asoft)_{i,j} &= \frac{e^{-(\mathcal{D})_{i,j}}}{\sum_{i}e^{-(\mathcal{D})_{i,j}}}.
\end{align}
A beta-binomial prior \cite{rad-tts} is then added to $\Asoft$ to promote elements on the near-diagonal line. Then, to extract discrete duration values, the Viterbi algorithm (also known as monotonic alignment search (MAS) \cite{glow-tts}) with diagonal constraints is applied to $\Asoft$. Specifically, we use dynamic programming to recursively solve $(Q)_{i, j}$, which is the score of a path that passes the $i$-th source feature at the $j$-th time step:
\begin{equation}
    (Q)_{i,j} = \max \bigl( (Q)_{i-1, j-1}, (Q)_{i, j-1} \bigr) + (\Asoft)_{i, j}. \label{eq:viterbi}
\end{equation}
Finally, the optimal path (i.e.  the hard alignment matrix $\Ahard$) can be traced by backtracking, and the ground truth duration sequence can thereafter be derived.

Two additional losses are added to the final objective in AAS-VC. First, a forward sum loss $\mathcal{L}_{\text{forward}}$ \cite{rad-tts} implemented in the form of connectionist temporal classification (CTC) \cite{ctc} maximizes the likelihood of the possible alignments. In our preliminary experiments, alignment learning fails without this loss. Second, as the $\mathcal{A}_{\text{hard}}$ generated by the Viterbi algorithm is non-differentialble, a KL loss $\mathcal{L}_{\text{KL}}$ minimizes the Kullback-Leibler divergence between $\Asoft$ and $\Ahard$. With $\alpha$ being a weight of the forward sum loss and the KL loss, the final objective of AAS-VC is then:
\begin{equation}
    \mathcal{L}_{\text{AAS-VC}}=\mathcal{L}_1 + \mathcal{L}_{\text{DP}} + \alpha(\mathcal{L}_{\text{forward}} + \mathcal{L}_{\text{KL}}),
\end{equation}

\subsection{Duration predictor}

While in \cite{cfs2-vc}, a simple L1 loss was used to train the \textit{deterministic} duration predictor, to tackle the one-to-many problem, flow-based \textit{probabilistic} duration predictors were proposed in the recent TTS literature to encourage diversity \cite{glow-tts, vits, jets}. At first sight, it is unclear whether (1) seq2seq VC needs this diversity, and (2) such duration predictors can be succeffsully trained on the comparatively limited VC data sizes, we will show that AAS-VC do benefit from such a probabilistic design.

\subsection{Source feature sequence length reduction}
\label{ssec:reduction}

In seq2seq VC, it is a common practice to reduce the length of the source feature sequence. While in \cite{VTN} a tiny strided CNN was used, \cite{e2e-s2s-vc} applied what we will refer to as ``pre encoder reduction (pre-er)'': given an input sequence $\{x_1, x_2, \cdots, x_{T}\} \in \mathbb{R}^{T\times d}$, the operation stacks $k$ adjacent vectors such that the resulting sequence becomes $\{[x_1, \cdots, x_k], \cdots, [x_{T-k+1}, x_{T}]\} \in \mathbb{R}^{(T/k) \times (d\times k)}$. $k$ is referred to as the \textit{source reduction factor}.


However, reducing the source feature sequence before the source encoder does not fully exploit the power of Conformers which are good at capturing long-range interactions. Instead, we propose to move the reduction operation after the encoder, which we refer to as ``post encoder reduction (post-er)''. As we will show later, this modification consistently improves the performance.

\subsection{Feature choice}
\label{ssec:feature}

Although mel-spectrogram is the de facto source feature choice in the seq2seq VC literature \cite{VTN, cfs2-vc}, many have also adopted phonetic posteriorgrams (PPGs)\footnote{In the modern era, the encoder outputs of neural ASR models is also referred to as ``PPG''.} as the source feature \cite{scent, scent-text}, which is not hard to justify as such a linguistically-rich feature should be a perfect fit for alignment learning in AR seq2seq VC.
In later sections, we will examine the generalization ability of PPG-based AAS-VC.
 
\section{Experimental evaluations}

\subsection{Experimental settings}

All experiments are carried out using the CMU ARCTIC database \cite{arctic}, where 1132 parallel recordings of US English speakers sampled at 16 kHz.
A male (bdl) and female source speaker (clb) as well as a male (rms) and female target speaker (slt) were used, resulting in four conversion pairs. Unless specified, all results are averaged over the four pairs. Each of the validation and evaluation sets contained 100 utterances, and the remainings were used as training data. 80-dimensional mel-spectrograms with 1024 FFT points and a 256
point frame shift were extracted. Speaker-dependent parallel waveGANs (PWGs) were used as the vocoder, and for comparison purposes, each PWG wes trained on the respective complete training sets of each target speaker, instead of training w.r.t. different training data sizes. The PPGs were extracted using an open-sourced Conformer ASR model \cite{ppg_url} trained on LibriSpeech 960 hrs \cite{librispeech}.

The implementation is open-sourced\footnote{\url{https://github.com/unilight/seq2seq-vc}} and we encourage interested readers to refer to it for details. The weight for $\mathcal{L}_{\text{forward}}$ and $\mathcal{L}_{\text{KL}}$ is set to 2, and the source reduction factor $k$ is set to 4. FS2-VC was chosen as the baseline, whose ground truth durations are distilled from the Voice Transformer Network (VTN) \cite{VTN}, an AR Transformer-based model. To show how FS2-VC depends on the quality of the AR model, we simulated two types of distilled durations of different quality. The first type is extracted from a VTN with pre-trained with the 24 hr LJSpeech dataset \cite{ljspeech} as described in \cite{VTN}, which we refer to as \textit{``PT''}. The second type is a VTN without pre-training, which we refer to as \textit{``No PT''}. It is expected that a VTN fails to train in a low-resource setting, thus the extracted durations fail to serve as good training targets for the FS2-VC model.

\subsection{Evaluation protocols}

Subjective \textit{naturalness} and speaker \textit{similarity} are the two main methods to evaluate VC systems.
In our experiments, listeners were asked to evaluate the naturalness on a five-point scale.
For conversion similarity, a natural target speech and a converted speech were presented, and listeners were asked to judge whether the two samples were produced by the same speaker on a four-point scale \cite{vcc2020}.
Both metrics are the higher the better.
We used crowd-sourcing to recruit 60 listeners to obtain 480 ratings per system.
Recordings of the target speakers (GT) and the analysis-synthesis samples from the vocoder (ANA-SYN) were also included to serve as the upper bound. Audio samples are available online\footnote{\url{https://unilight.github.io/Publication-Demos/publications/aas-vc/index.html}}.

Mel cesptral distortion (MCD) is a general objective metric in VC that well correlates with human perception in a homogeneous setting. Character error rates (CERs) from an off-the-shelf ASR model is used to access intelligibility. To measure the main ability of seq2seq VC to model prosody, F0 linear correlation (F0CORR) and absolute duration difference (DDUR) are also reported. In addition, we introduce a new metric called the duration variance (DVAR), which is defined as the variance of all frame-level duration predictor outputs. A zero DVAR indicates that the duration predictor only learns a linear duration mapping, which should be avoided in seq2seq models. A larger DVAR value, on the other hand, encourages diversity. Note this only applies to non-AR models. Except for F0CORR and DVAR, the scores of MCD, CER, DDUR and  are the smaller the better.

\subsection{Justifying model design choices}

In this subsection we verify the effectiveness of the model design choices by using 932 training utterances ($\sim 1$ hr). Results are shown in Table~\ref{tab:ablation}, and row (a) is the proposed combination.

\begin{table}[t]
    \footnotesize
	\centering
	\caption{Objective comparison of model design choices using 932 training utterances. ``DP'' stands for ``duraiton predictor'', and ``Det.'', ``Sto'' stand for ``deterministic'', ``stochastis'', respectively.}
	
	\centering
        \begin{tabular}{c c c c | c c c }
		\toprule
            & \makecell{DP\\type} & \makecell{Source\\reduction}  & \makecell{Source\\feature} & MCD $\downarrow$ & CER $\downarrow$ & DVAR $\uparrow$ \\
            \midrule
            (a) & Sto. & post-er & mel & 6.35 & 3.4 & 0.631 \\
            \midrule
            (b) & Det. & post-er & mel & 6.38 & 3.0 & 0.303 \\
            (c) & Sto. & CNN & mel & 6.44 & 5.2 & 0.504 \\
            (d) & Sto. & pre-er & mel & 6.40 & 5.8 & 0.708 \\
            (e) & Sto. & post-er & PPG & 6.23 & 1.7 & 0.389 \\
            \bottomrule
	\end{tabular}
	\label{tab:ablation}
\end{table}


\noindent{\textbf{Duration predictor}:} By comparing rows (a) and (b) in Table~\ref{tab:ablation}, we can see the deterministic duration predictor yields similar CER and MCD values but a much lower DVAR. This shows how a stochastic duration predictor encourages diversity in duration modeling.


\noindent{\textbf{Ways of reducing source length}:} We compare using a strided CNN (row (c)) or simply stacking adjacent frames before the Conformer encoder (row (d)) with post-er (row (a)). The effectiveness of post-er is justified by the better MCD and CER values compared to those of CNN and pre-er. Although pre-er has a higher DVAR, we suspect that such diversity could be a potential cause of the increased CER. We therefore choose to use post-er in the subsequent experiments.


\noindent{\textbf{Feature choice}:} Comparing rows (a) and (e) in Table~\ref{tab:ablation}, using PPG yields a much lower CER and a better MCD value. This is a natural outcome since for the encoder, extracting compact linguistic representations from PPG is much easier compared to disentangling contents from the many other factors in mel-spectrograms. As for the low DVAR, we suspect it was caused by the smaller hop size used in the ASR model. We would like to note that, as it is difficult to develop a customized ASR model for atypical speech like accented, dysarthric and EL speech, the default feature choice in AAS-VC is still mel-spectrogram.

\begin{figure}[t]
	\centering
	\includegraphics[width=\linewidth]{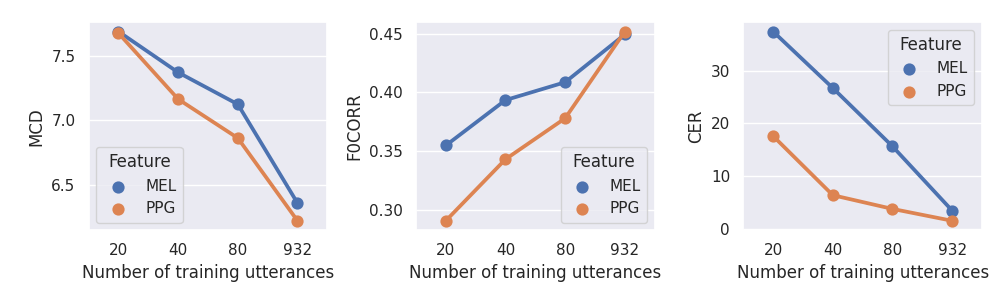}
	\caption{\label{fig:mel-vs-ppg}Objective comparison of using mel-spectrograms and PPG as the source feature in various training data sizes.}	
 
\end{figure}

To further understand the pros and cons of using PPG, we varied the training data size, and the objective results are shown in Figure~\ref{fig:mel-vs-ppg}. When the training data size decreases, in terms of intelligibility, the advantage of using PPG becomes more obvious, but on the other hand, the F0CORR decreases more rapidly than that of using mel-spectrograms. This is possibly due to the very little to none f0 information left in the ``overly disentangled'' PPG . As a result, the model's heavy workload to reconstruct f0 ``from scratch'' results in almost identical MCD values of using mel-spectrogram and PPG when the number of training utterances falls to 20.

\vspace{-5pt}
\subsection{The dependency of FS2-VC on quality of distilled durations}
\vspace{-5pt}

\begin{figure}[t]
	\centering
	\includegraphics[width=\linewidth]{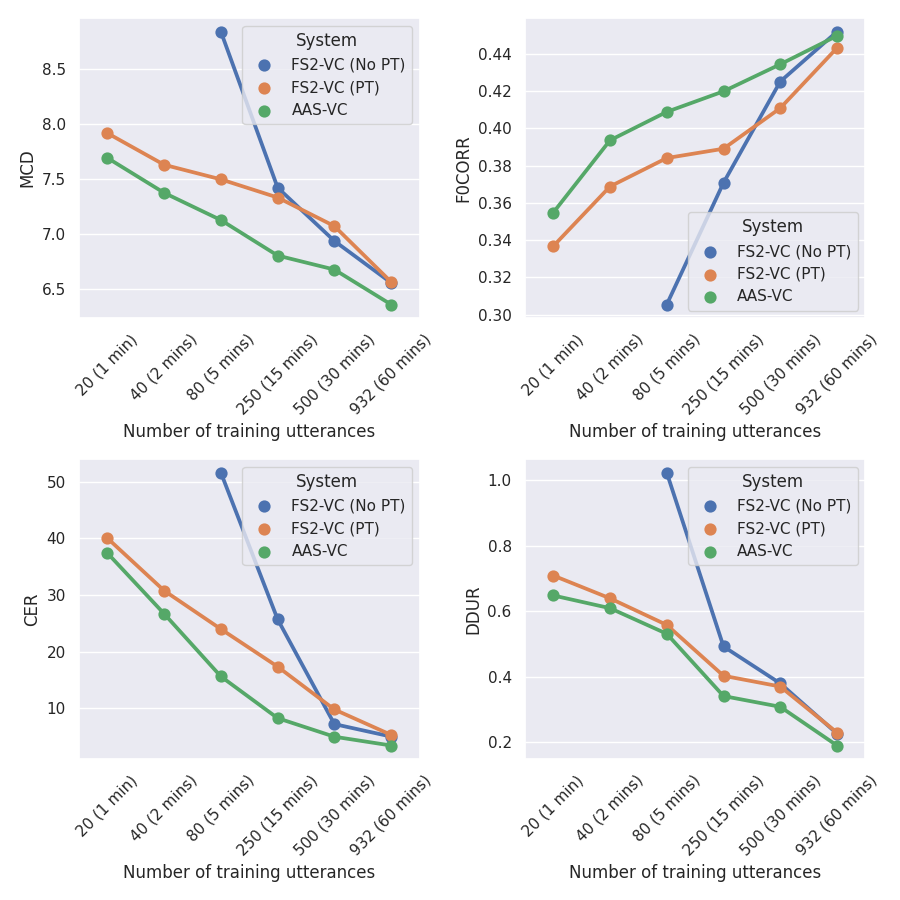}
	\caption{\label{fig:main}Objective comparison of different seq2seq VC systems in various training data sizes. Note that training of FS2-VC (No PT) on 40 and 20 training utterances failed due to unusable distilled durations.
 \vspace{-5pt}
 }	
\end{figure}

\begin{table}[t]
    \footnotesize
	\centering
	\caption{Subjective evaluation results of different seq2seq VC systems in various training data sizes. Bold face indicates the top performing system(s) in each column (multiple bolded systems indicate that there is no significant difference in between).}
        \footnotesize
	
	\centering

        \begin{tabular}{l | c c | c c }
		\toprule
            \multirow{3}{*}[-1pt]{System} & \multicolumn{4}{c}{Number of training utterances (duration)} \\
            \cmidrule{2-5}
            & \multicolumn{2}{c|}{80 (5 mins)} & \multicolumn{2}{c}{932 (1 hr)}   \\
            & Nat. $\uparrow$ & Sim. $\uparrow$ & Nat. $\uparrow$ & Sim. $\uparrow$ \\
            \midrule
            GT & \multicolumn{2}{c|}{-} & 4.42$\pm$0.09 & 81\%$\pm$3\% \\
            ANA-SYN & \multicolumn{2}{c|}{-} & 3.72$\pm$0.12 & 76\%$\pm$4\% \\
            \cmidrule{1-5}
            FS2-VC (No PT) & 1.84$\pm$0.11 & 41\%$\pm$4\% & \textbf{3.62$\pm$0.11} & \textbf{65\%$\pm$4\%} \\
            FS2-VC (PT) & 2.34$\pm$0.11 & \textbf{50\%$\pm$4\%} & \textbf{3.52$\pm$0.11} & \textbf{67\%$\pm$4\%} \\
            AAS-VC & \textbf{2.80$\pm$0.12} & \textbf{51\%$\pm$4\%} & 3.47$\pm$0.12 & \textbf{65\%$\pm$4\%} \\
            \bottomrule
	\end{tabular}
 
	\label{tab:subjective}
\end{table}

We then compare FS2-VC trained on different distilled durations. From Figure~\ref{fig:main}, it can be clearly observed that while FS2-VC (No PT) and FS2-VC (PT) performed similarly with 1 hr of training data, the latter can generalize better in low-resource settings. The same tendency is also true in the subjective evaluation results presented in Table~\ref{tab:subjective}, where FS2-VC (PT) outperformed FS2-VC (No PT) significantly in the 5 mins training set setting.

\subsection{Comparison of FS2-VC and AAS-VC}

Next we compare AAS-VC with FS2-VC w.r.t. different training data sizes. Figure~\ref{fig:main} shows that AAS-VC consistently outperformed the baselines in most objective metrics w.r.t. all training data sizes. Table~\ref{tab:subjective} shows that in the 5 mins setting, AAS-VC has a superior subjective performance in both naturalness and similarity, and there was no significant difference in terms of similarity in the 1 hr setting. As for naturalness, the p-values between AAS-VC and FS2-VC (PT), FS2-VC (No PT) are 0.015, 0.553, respectively. The overall result is considered remarkable since AAS-VC can achieve the same level of generalization ability while FS2-VC (PT) benefitted from pre-training on a large TTS dataset. Such results also justified that the AAS is an effective inductive bias to reduce the data requirement. 





\vspace{-10pt}
\section{Conclusion}
In this work, we investigate and improve the generalization ability of non-AR seq2seq VC models. We first show that FS2-VC depends heavily on the quality of the distilled durations from the AR model. We then present AAS-VC, a non-AR seq2seq VC model with AAS that can derive directly from the training utterance pairs the alignments as well as the ground truth durations for training. Experimental evaluation results show that AAS-VC can generalize to small training datasets better than FS2-VC. Future works include extensions to unseen source speakers, as well as applying AAS-VC to different VC sub-applications.

\noindent{\textbf{Acknowledgment:}} This work was partly supported by JSPS KAKENHI Grant Number 21J20920 and JST CREST Grant Number JPMJCR19A3.

\bibliographystyle{IEEEbib}
\bibliography{refs}

\end{document}